\documentclass[12pt]{article}
\usepackage{graphicx}

\newcommand\pubnumber{SNSN-323-63}
\newcommand\pubdate{\today}
\newcommand{\be}{\begin{equation}}
\newcommand{\ee}{\end{equation}}

\newcommand{\msb}{{\overline{\rm MS}}}

\def\napoli{Department of Physics,
The Ohio State University,
Columbus, OH 43210, USA.}
\def\Title#1{\begin{center} {\Large #1 } \end{center}}
\def\Author#1{\begin{center}{ \sc #1} \end{center}}
\def\Address#1{\begin{center}{ \it #1} \end{center}}

\newcommand\pubblock{\rightline{\begin{tabular}{l} \pubnumber\\
         \pubdate  \end{tabular}}}
\newenvironment{Abstract}{\begin{quotation}  }{\end{quotation}}
\newenvironment{Presented}{\begin{quotation} \begin{center} 
             \end{center}\bigskip 
      \begin{center}\begin{large}}{\end{large}\end{center} \end{quotation}}
\def\Acknowledgements{\bigskip  \bigskip \begin{center} \begin{large}
             \bf ACKNOWLEDGEMENTS \end{large}\end{center}}
\def\beq{\begin{equation}}
\def\eeq#1{\label{#1}\end{equation}}
\def\eeqn{\end{equation}}

\def\beqa{\begin{eqnarray}}
\def\eeqa#1{\label{#1}\end{eqnarray}}
\def\eeqan{\end{eqnarray}}

\let\bar=\overbar

\def\Dslash{\not{\hbox{\kern-4pt $D$}}}
\def\dslash{\not{\hbox{\kern-2pt $\del$}}}

\def\ee{e^+e^-}

\def\msb{{\bar{\ssstyle M \kern -1pt S}}}

\begin{document}
\begin{titlepage}
\pubblock

\vfill
\Title{Progress in Lattice QCD Relevant for Flavor Physics}
\vfill
\Author{ Junko Shigemitsu}
\Address{\napoli}
\vfill
\begin{Abstract}
Recent Lattice QCD results relevant for Kaon, Charm and 
$B$ Physics are summarized. There is general agreement among 
calculations using a wide range of different lattice actions.
This bolsters confidence in the lattice results and in their quoted 
errors. One notes considerable progress  
since CKM2008 in reducing lattice errors with some quantities now being 
calculated at the subpercent to a few percent level accuracy. 
 Much work remains, however, 
and further improvements can be expected in the coming years.
\end{Abstract}
\vfill
\begin{Presented}
Proceedings of \underline{CKM2010}\\
The 6th International Workshop on the \\
CKM Unitarity Triangle\\
University of Warwick, UK, 6-10 September 2010
\end{Presented}
\vfill
\end{titlepage}
%\def\thefootnote{\fnsymbol{footnote}}
%$\setcounter{footnote}{0}

\section{Introduction}
A major goal of Lattice QCD is to carry out the theoretical 
calculations that are necessary and relevant to the Flavor Physics 
Program in Particle Physics. There has been huge progress 
in recent years in working towards this goal.  The most accurate lattice 
calculations now typically have errors at the  
$ 0.5 \, - \, 4$\% level in Kaon physics ($f_+^{K \rightarrow \pi}(0)$, 
$f_K/f_\pi$, $B_K$, ...), at the $ 1 \, - \, 10$\% level in Charm physics 
($f_{D_s}/f_D$, $f_D$, $f_{D_s}$, $f_+^{D \rightarrow K}(0)$, 
$f_+^{D \rightarrow \pi}(0)$, ...) and at the $ 2 \, - \, 10$\% level
in $B$ Physics (${\cal F}(1)$, $f_{B_s} / f_B$, $f_B$, $f_{B_s}$, 
$\xi$, $f_+^{B \rightarrow \pi}(q^2)$, ....).  It is particularly noteworthy 
that sub-percent accuracy has been achieved in Kaon physics and 
that the ability to simulate very light quarks has improved dramatically, with the 
pion mass approaching the physical value ever more closely. 
Furthermore, recent years have witnessed significant improvements in our 
ability to simulate charm quarks accurately on the lattice.  Much work remains, 
however in Lattice Flavor Physics especially in $B$ physics. 
In this review talk, I would like to present a brief summary of recent 
results and plans for future improvements.  Due to page restrictions, 
several topics and all details will have to be omitted.

\section{Kaon Physics}

%%%%%%%%%%%%%%%%%%%%%%%%%%%%%%%%%%%%%%%%%%%%%%%%%%%%%%%%%%%%%%%%%%%%%%%%%
%%
%%   use this format to include a LaTeX table  into your paper
%%
%\begin{table}[t]
\begin{table}
\begin{center}
\begin{tabular}{|c|c|c|c|c|}  
\hline
Collaboration (yr) & $f_+(0)$ & $N_f$ & \# lattice spacings & action \\ 
\hline
RBC/UKQCD (07) \cite{rbcfplus07}   &0.964$(5)$   & 2 + 1 &  1 & domain wall \\
RBC/UKQCD (10) \cite{rbcfplus10} & 0.960$\left (^{+5}_{-6} \right )$   & 2 + 1 &  1 &
 domain wall\\
ETMC (09) \cite{etmcfplus09} & 0.956(8)   &  2  &  3  & twisted mass  \\
\hline 
\end{tabular}
\caption{Semileptonic Kaon decay form factor $f_+^{K \rightarrow \pi}(0)$}
\label{tab:blood}
\end{center}
\end{table}
%%%%%%%%%%%%%%%%%%%%%%%%%%%%%%%%%%%%%%%%%%%%%%%%%%%%%%%%%%%%%%%%%%%%%%%%%%%
A recent global analysis by FlaviaNet demonstrates the precision with 
which $K_{l3}$ and $K_{l2}$ decays are testing the Standard Model (SM). 
For instance, tests of first row unitarity now stand at 
\cite{flavianet,pdg2010},
$$
\Delta_{CKM} = |V_{ud}|^2 + |V_{us}|^2 + |V_{ub}|^2 - 1 = 0.0001(6).
$$
Lattice QCD input is crucial for determining $|V_{us}|$ \cite{flag}.
 It provides the 
form factor $f_+^{K \rightarrow \pi}(0)$ for analysis of 
 $K_{l3}$ decays and the ratio of decay constants $f_K / f_\pi$  
for the $K_{l2}$ analysis. Table 1 shows recent lattice results for 
$f_+^{K \rightarrow \pi}(0)$. The RBC/UKQCD collaboration uses domain wall 
fermions \cite{rbcfplus07,rbcfplus10}.
 Their 2010 update works directly at $q^2 = 0$ and employs 
a more sophisticated chiral perturbation theory (ChPT) analysis 
than in their earlier work. 
In the ETMC calculations based on twisted mass quarks \cite{etmcfplus09}, 
strange sea quark contributions were not included 
directly but their effects were estimated and included through NLO in ChPT.
Fig.1 summarizes lattice results for the ratio $f_K/f_\pi$ from 
many lattice collaborations using a variety of lattice 
quark actions \cite{milcfkfpi10,milcfkfpi09,aubinfkfpi,hpqcdfkfpi,bmwfkfpi,
rbcfkfpi,etmcfkfpi}.
 An average by the FlaviaNet group, $f_K/f_\pi = 1.193(6)$ \cite{flavianet}
 is also shown.  

\noindent
There has also been progress on the $K^0 - 
\overline{K}^0$ mixing parameter $B_K$.  Fig.2 shows a summary and 
one sees excellent agreement between results from 
different lattice actions \cite{aubinbk,rbcbk07,rbcbk10,etmcbk,sbwbk}. 
An average of a subset of the data by Laiho-Lunghi-VandeWater 
(LLV), $\hat{B}_K = 0.725(26)$ \cite{llv}, is included in the plot.

\parbox{0.5\hsize}{
\includegraphics*[height=11.cm,angle=270]{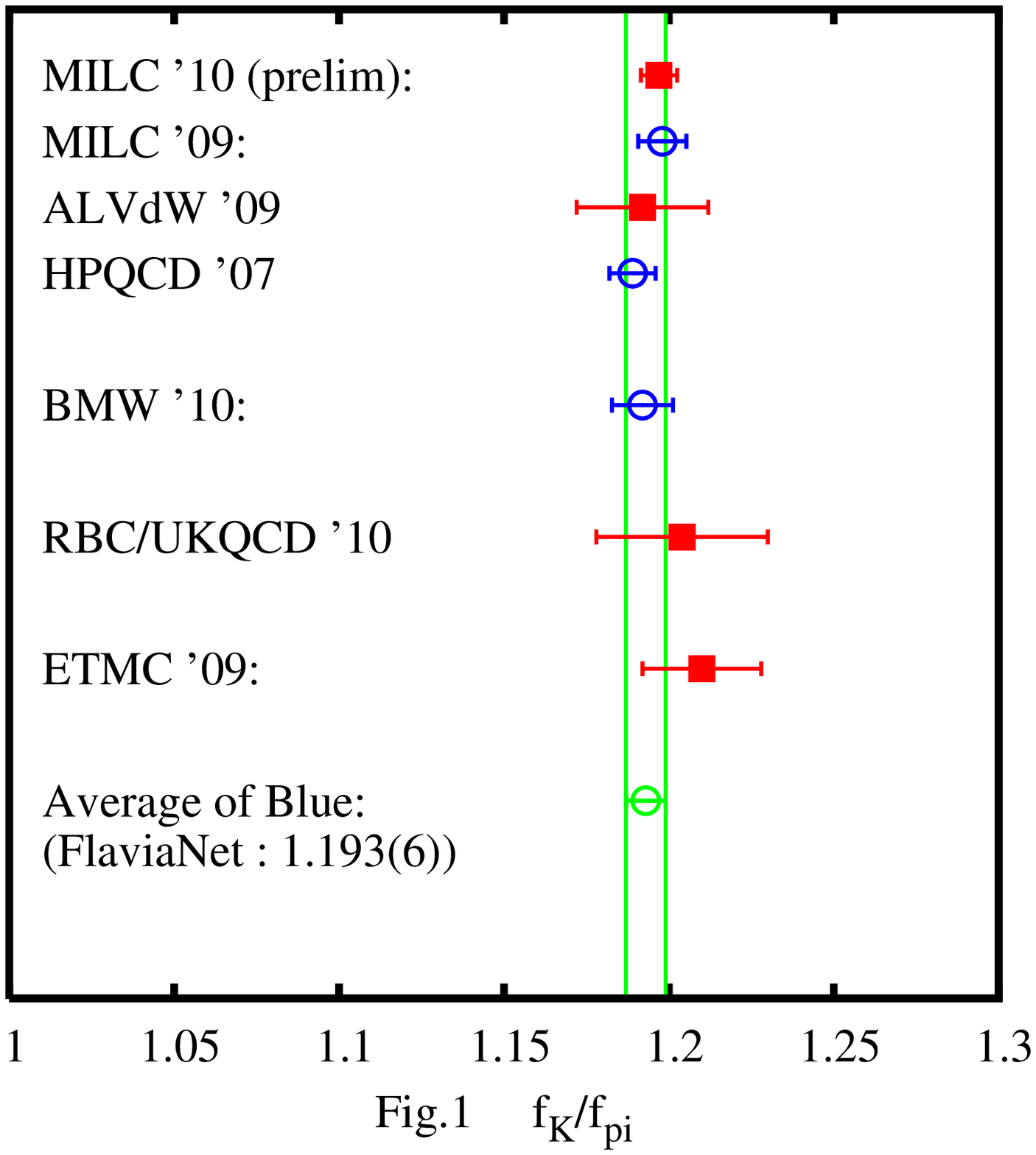}
}
\parbox{0.5\hsize}{
\includegraphics*[height=11.cm,angle=270]{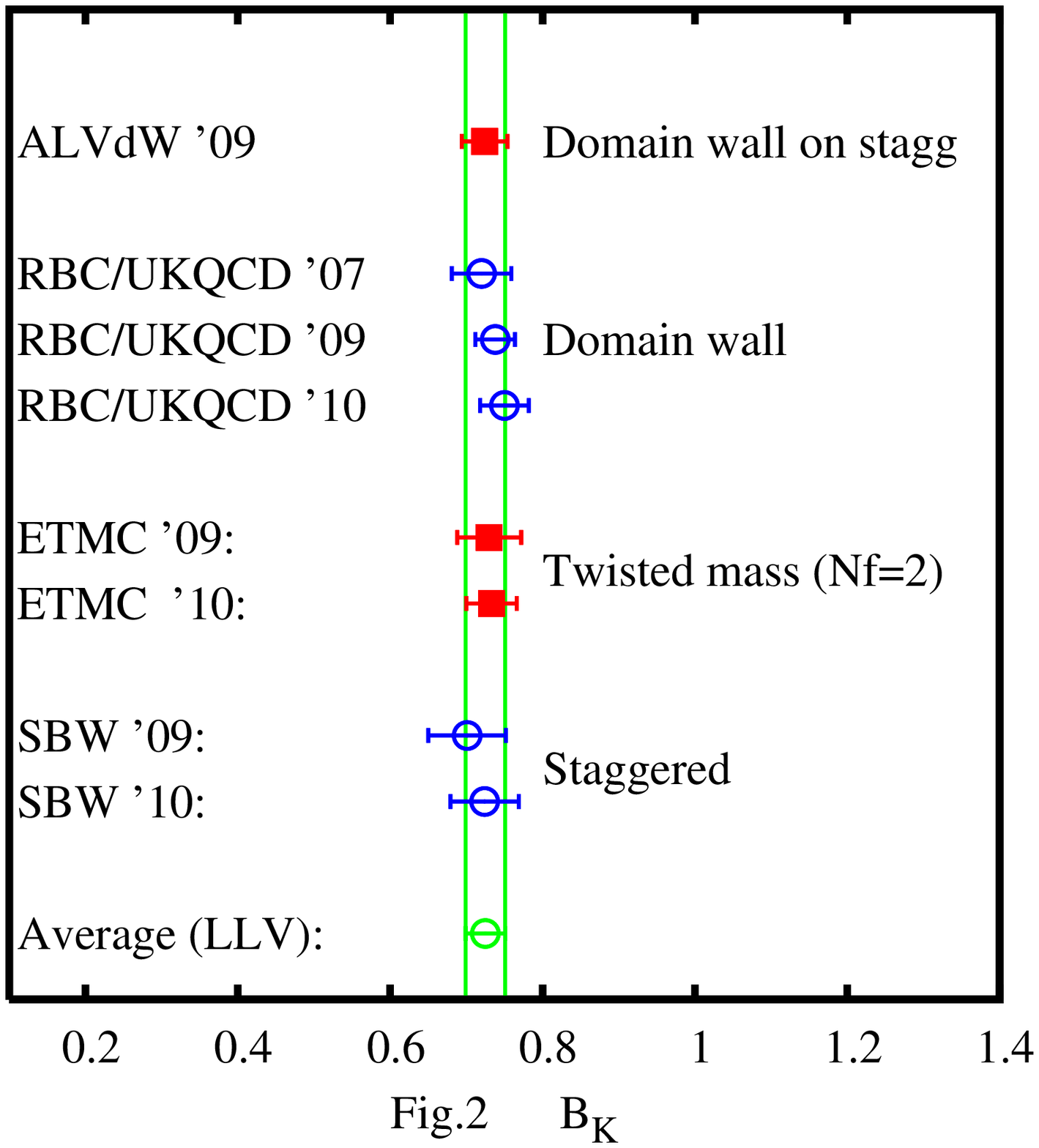}
}

\vspace{.1in}
\noindent
The next frontier in Kaon physics is $K \rightarrow \pi \pi$ in the 
$\Delta I = 3/2$ and $\Delta I = 1/2$ channels \cite{chris}. 
  Progress has been made 
recently  by two groups using 
different approaches.  RBC/UKQCD \cite{rbckpipi} uses a ``direct'' Lellouch-Luescher 
finite volume method. Coumbe-Laiho-Lightman-VandeWater \cite{laihokpipi} first work at an 
unphysical kinematic point, $2 M_\pi = M_K$ in order to evade the 
Maiani-Testa No-Go Theorem that afflicts Euclidean lattice simulations 
of this process. They then use ChPT to get back to physical pions. 
 At least for the $\Delta I = 3/2$ case, both methods appear 
feasible with errors at the 15 - 20\% level.
%%%%%%%%%%%%%%%%%%%%%%%%%%%%%%%%%%%%%%%%%%%%%%%%%%%%%%%%%%%%%%%%%%%%%%%%%
%%
%%   use this format to include an .eps figure into your paper
%%
%\begin{figure}[htb]
%\begin{figure}
%\centering
%\includegraphics[height=4in,angle=270]{fkfpi.ps}
%\caption{}
%\end{figure}
%%%%%%%%%%%%%%%%%%%%%%%%%%%%%%%%%%%%%%%%%%%%%%%%%%%%%%%%%%%%%%%%%%%%%%%%%%%

\section{Charm Physics}
Studies of leptonic decays of the $D$ and in particular of the 
$D_s$ mesons have attracted considerable attention in the last couple of years. 
Experiment and theory are compared against each other via the 
decay constants $f_D$ and $f_{D_s}$. Table 2 summarizes lattice 
calculations of the decay constants as of summer 2010. 
\begin{table}
\begin{center}
\begin{tabular}{|c|c|c|c|}
\hline
Collaboration  &  $f_D$ (MeV)
  &  $f_{D_s}$ (MeV) &  $f_{D_s} /f_D$ \\
\hline
 Fermi/MILC '05 \cite{fermifd05}  &  201 $\pm$ 17  &  249 $\pm$ 16   &
 1.24 $\pm$ 0.07  \\
\hline
 Fermi/MILC '10 \cite{fermifd10} &  220 $\pm$ 9 $\pm$ 5  &
 261 $\pm$ 8 $\pm$ 5   &  1.19 $\pm$ 0.01  \\
( preliminary) & & &  $ \quad \; \; \; \; \pm$ 0.02\\
\hline \hline
HPQCD '07 \cite{hpqcdfkfpi} & 207 $\pm$ 4 &  241 $\pm$ 3 & 1.164 $\pm$ 0.011 \\
\hline
HPQCD '10 \cite{hpqcdfd10}  &  &  248.0 $\pm$ 2.5 &  \\
\hline
\hline
 ETMC '09  ($N_f = 2$) \cite{etmcfkfpi}  &  197 $\pm$ 9
  &   244 $\pm$ 8  &  1.24 $\pm$ 0.03 \\
\hline
\end{tabular}
\caption{Decay constants $f_D$ and $f_{D_s}$}
\end{center}
\end{table}
\noindent
Experimental determinations of 
decay constants start from the branching fractions for the leptonic decays 
of charged $D_q$ mesons ${\cal B}(D_q \rightarrow l \nu) $,
 corrected for electromagnetic 
effects ($q = d$ or $s$). 
In the Standard Model these branching fractions depend on 
the combinations $f_{D_q} * |V_{cq}|$.  Using estimates of the CKM matrix 
elements $|V_{cq}|$ based on unitarity one then has,
$f_{D_q}^{exp.}$ \cite{cleofd,hfagfds}.

\parbox{0.5\hsize}{
\includegraphics*[height=11.cm,angle=270]{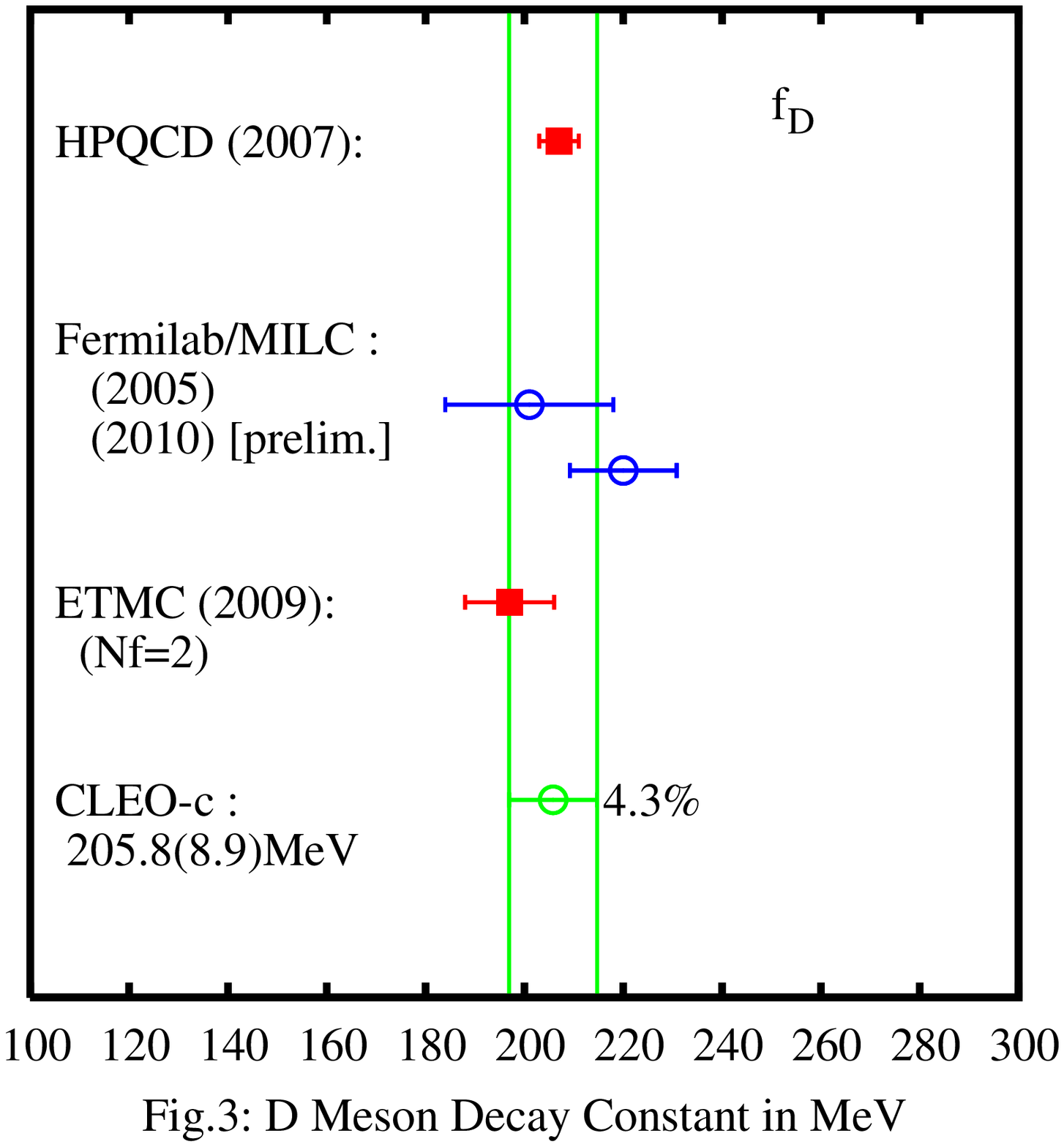}
}
\parbox{0.5\hsize}{
\includegraphics*[height=11.cm,angle=270]{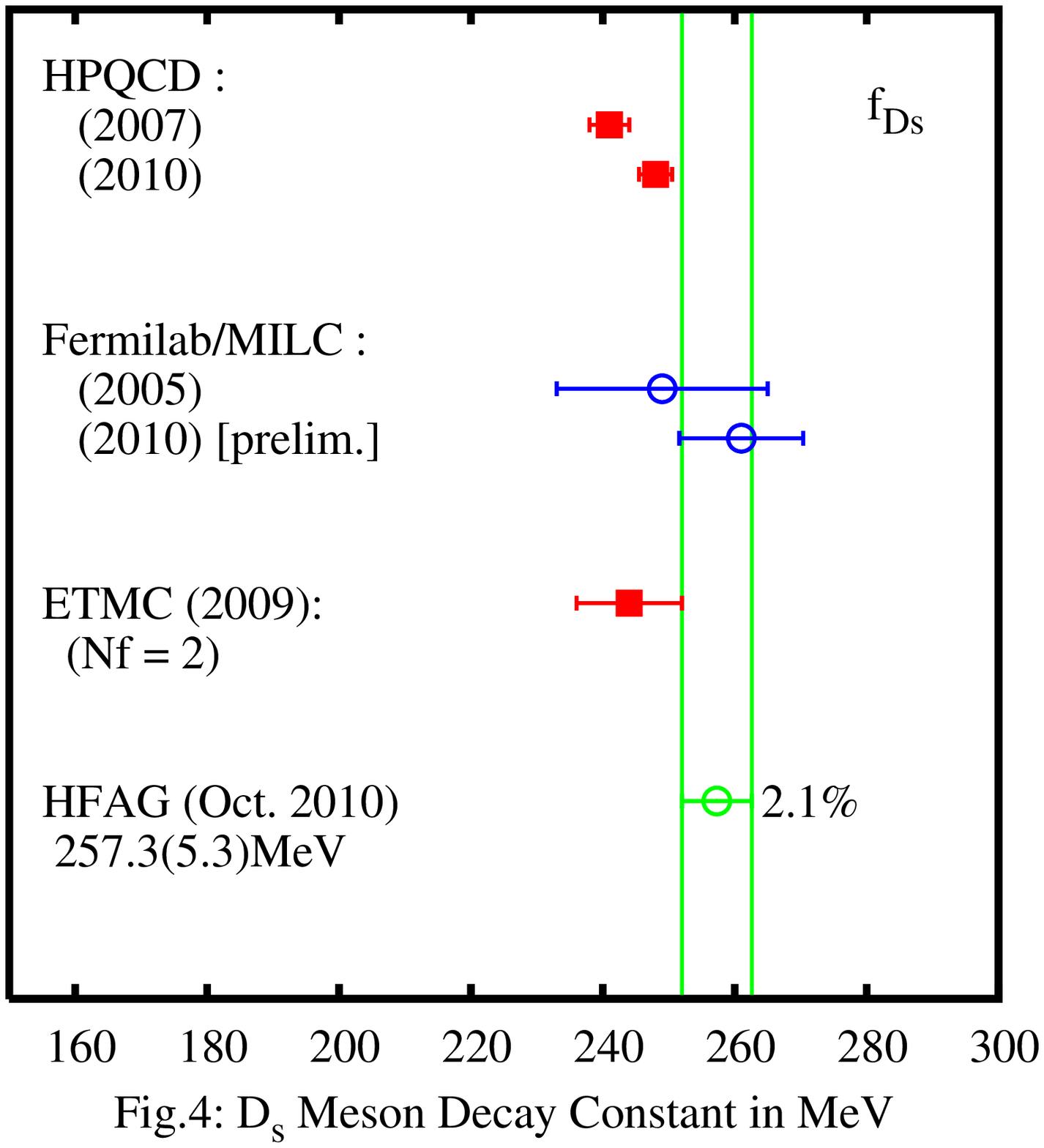}
}

\vspace{.1in}
\noindent
Comparisons between experiment and the Lattice QCD results of Table 2 are 
shown in Figs.3 and 4. For $f_D$ agreement is very good. For $f_{D_s}$ 
the most accurate 2010 HPQCD result and the HFAG October 2010 experimental 
average \cite{hfagfds} differ by about $1.6 \sigma$.  It will be interesting to see 
how things develop in the future when more lattice calculations from other groups 
and experimental results reach the $\sim 1$\% level of accuracy.

\vspace{.1in}
\noindent
Significant progress can also be reported on in studies of $D$ meson 
 semileptonic decays on the lattice.  Here the 
goal is to obtain the form factors $f_+^{D \rightarrow K}(q^2)$ 
and $f_+^{D \rightarrow \pi}(q^2)$ and combine this with 
experimental measurements of $|V_{cq}| * f_+(0)$ to determine $|V_{cd}|$ 
and $|V_{cs}|$. A very recent result by the HPQCD collaboration \cite{hpqcdf0} 
for the $D \rightarrow K$ form factor at $q^2 = 0$ is given below 
together with $|V_{cs}|$ extracted using experimental input from 
CLEO-c \cite{cleof0} and BaBar \cite{barbf0}.   Fig.5 plots the HPQCD result 
together with earlier theory calculations, both from the lattice \cite{fermif0} and 
using lightcone sum rules methods \cite{sumrf0}.  The theory error has been 
reduced by about a factor of four by the 2010 HPQCD calculations.  
The main reasons for this improvement come from the use of a 
highly improved lattice action (HISQ) for both the charm and the 
lighter quarks, the exploitation of hadronic matrix elements that 
do not require any renormalization and novel approaches to fitting and 
to chiral/continuum extrapolations of lattice data.

\parbox{0.5\hsize}{
\includegraphics*[height=10.cm,angle=270]{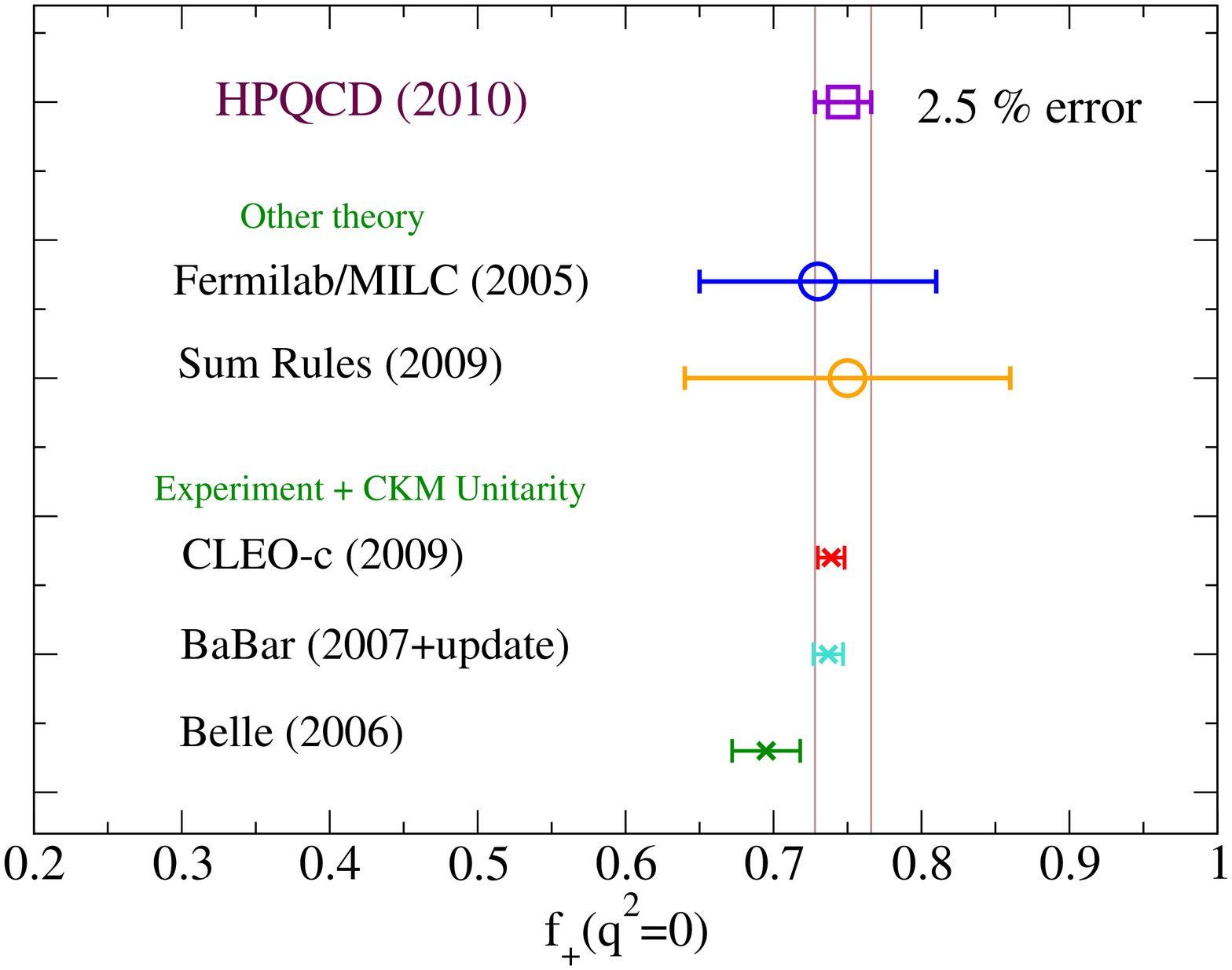}
}
\parbox{0.5\hsize}{

\vspace{.2in}

\hspace*{1.in}  HPQCD 2010 \cite{hpqcdf0} : 

\vspace{.2in}

$ \qquad \qquad \qquad f_+^{D \rightarrow K}(0) = 0.747(19)$

\vspace{.4in}
$ \qquad \qquad \qquad \Longrightarrow$

$ \qquad \qquad \qquad |V_{cs}| = 0.961(11)(24)$\\
\hspace*{1.in} 1$^{st}$ error from exper.\\
\hspace*{1.in} 2$^{nd}$ error from theory.

\vspace{.4in}

}
 \hspace*{2.in} Fig.5

\vspace{.1in}
\noindent
Fig.5 also includes experimental determinations of $f^{D \rightarrow K}_+(0)$
 \cite{cleof0,barbf0,bellef0}. 
These were obtained by using the unitarity value for $|V_{cs}|$.  One sees 
that the CLEO-c and BaBar results are already very precise, with errors 
at the $\sim 1$\% level a factor of two smaller than for the best lattice 
calculations.  So, inspite of the recent impressive progress on the
lattice side, further improvements are called for.

\noindent
The HPQCD collaboration has used its new $|V_{cs}|$ to test CKM unitarity 
\cite{hpqcdf0}.   
Using PDG values $|V_{cd}| = 0.230(11)$ and $|V_{cb}| = 0.0406(13)$ one 
finds for 2nd row unitarity,
$$|V_{cd}|^2 + |V_{cs}|^2 + |V_{cb}|^2 = 0.978(50).$$
Similarly for the 2nd column with $|V_{us}| = 0.2252(9)$ and $|V_{ts}|= 0.0387(21)$ 
one finds
$$|V_{us}|^2 + |V_{cs}|^2 + |V_{ts}|^2 = 0.976(50).$$
These unitarity tests are more accurate than the current PDG 2010 results \cite{pdg2010}  
of 1.101(74) and 1.099(74) for the 2nd row or 2nd column respectively.

\vspace{.1in} 
\noindent
The Fermilab/MILC collaborations are updating their 2005 $D$
 semileptonic form factor  
calculations with improved statistics and chiral/continuum extrapolations 
using many more lattice spacings \cite{fermif010}. Preliminary results for the 
scaled $f_+(q^2) / f_+(0.15)$ and a comparison with CLEO-c data are
given in Fig.6.  Fig.7 shows preliminary results by the ETM collaboration \cite{etmcf0}. 
In both cases the lattice calculations reproduce the shape of the form factor 
as a function of $q^2$ in good agreement with experiment. Fermilab/MILC and ETMC 
are working towards final results for $|V_{cs}|$ and $|V_{cd}|$.  HPQCD 
is also repeating its $D \rightarrow K$ calculations for the $D \rightarrow \pi$ case.

\begin{center}
\includegraphics*[height=12.cm,angle=270]{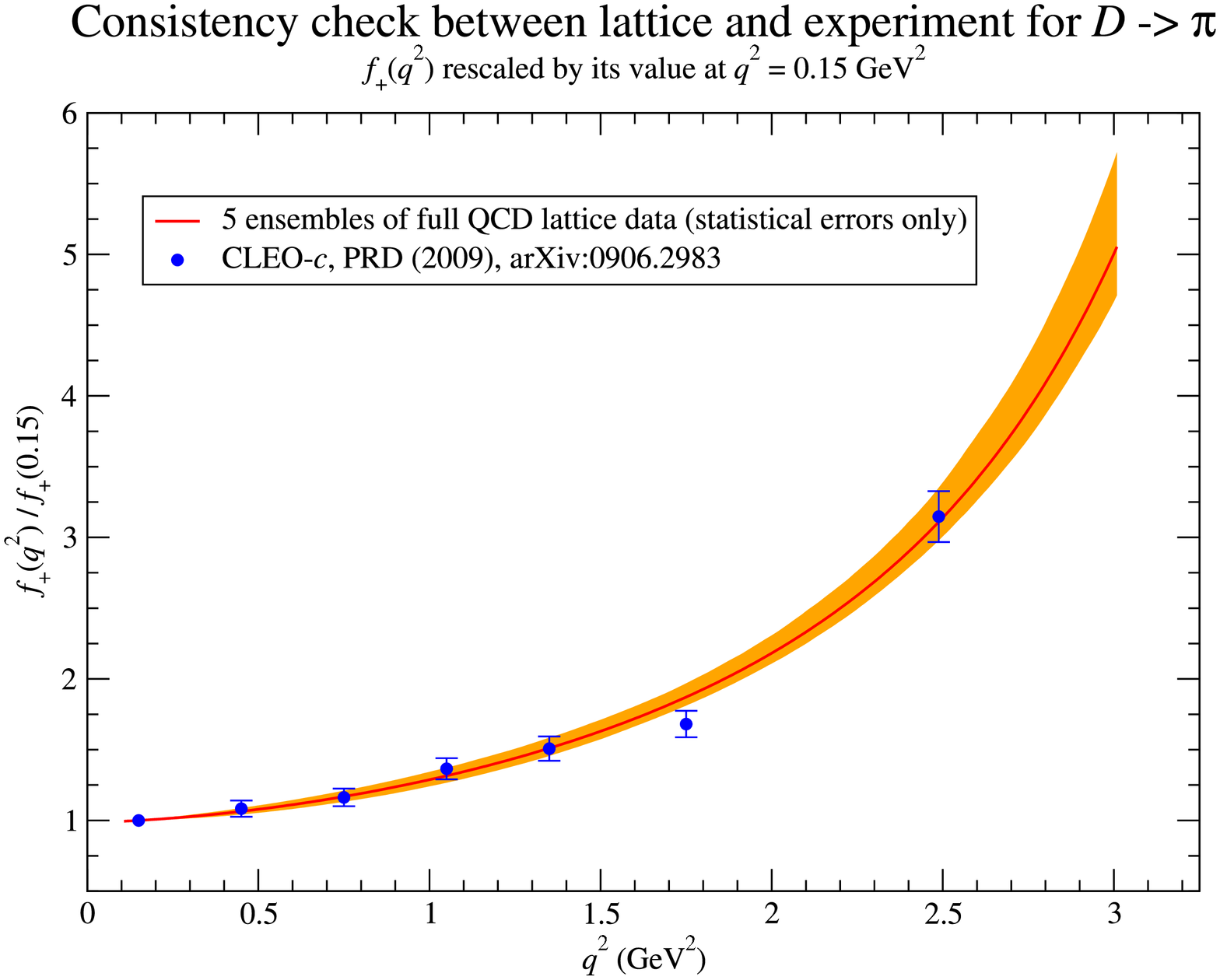}
\end{center}
\hspace*{2.5in} Fig.6:  Fermilab/MILC \cite{fermif010}

\vspace{.3in}
\parbox{0.5\hsize}{
\includegraphics*[height=5.5cm]{ETMC-Di_Vita-D_to_K_Form_Factors.eps}
}
\parbox{0.5\hsize}{
\includegraphics*[height=5.5cm]{ETMC-Di_Vita-D_to_pi_Form_Factors.eps}
}
\hspace*{.5in} Fig.7: ETMC results for $D \rightarrow K$ (left) 
and $D \rightarrow \pi$ (right) form factors \cite{etmcf0}

\section{$B$ Physics}
Since CKM2008 there have been updates of 
the $B \rightarrow D^*$ form factor at zero recoil ${\cal F}(1)$, 
new results for $B - \overline{B}$ mixing and several calculations/updates 
of the decay constants $f_B$ and $f_{B_s}$ \cite{heitger}. 
 Fermilab/MILC's results for 
${\cal F}(1)$ are given below \cite{fermif109,fermif110}.  Fig.8 shows their most recent 
chiral/continuum extrapolations from 4 lattice spacings.  

\vspace{.2in}

\parbox{0.6\hsize}{
\includegraphics*[height=9.cm]{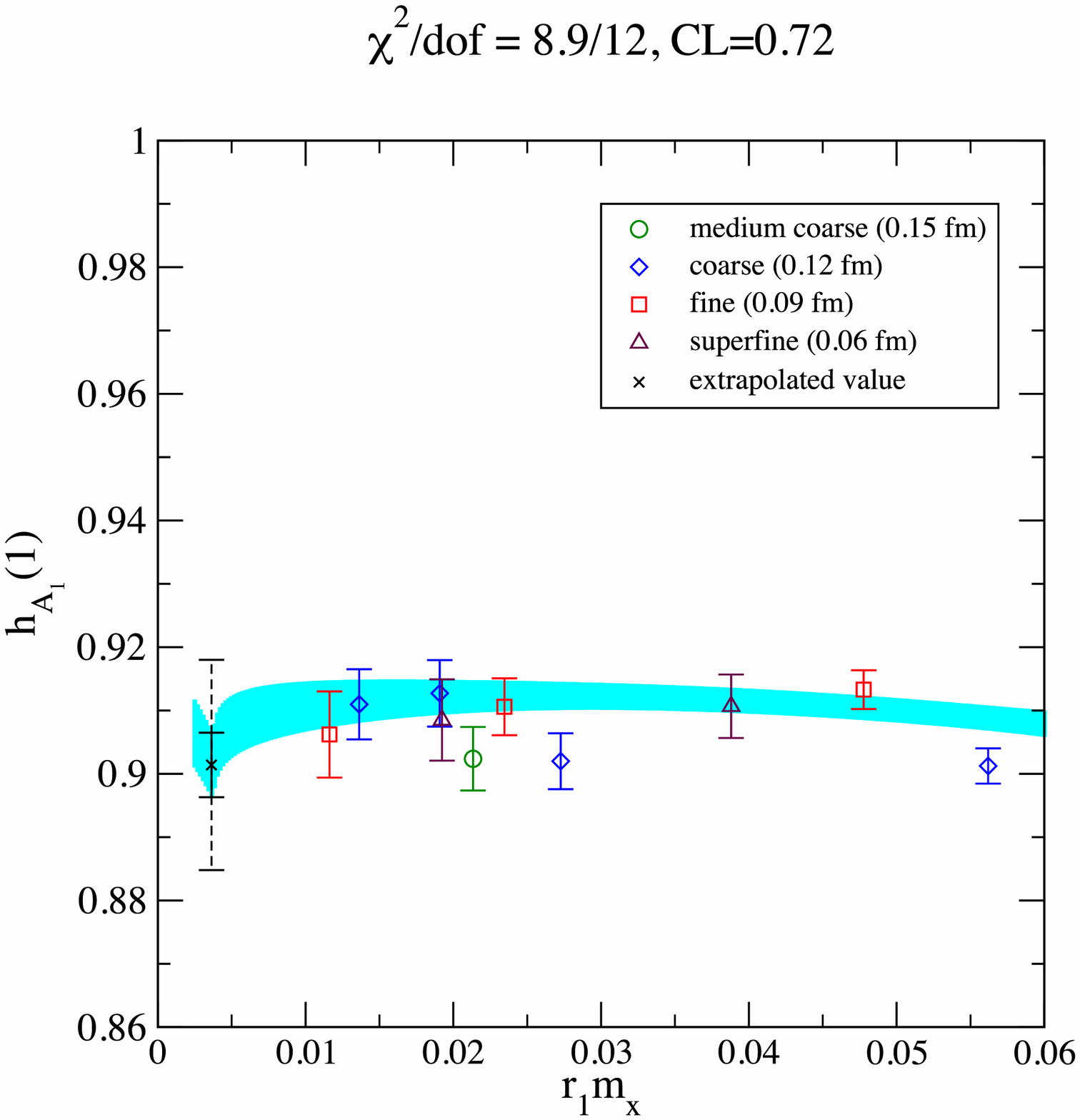}
}
\parbox{0.4\hsize}{

\vspace{.3in}

  Fermilab/MILC : 

\vspace{.1in}
\underline{2008} \cite{fermif109}

${\cal F}(1) = 0.921(13)(20)$ \\
Lattice error of 2.6\%

\vspace{.1in}
\underline{2010} (preliminary) \cite{fermif110}

${\cal F}(1) = 0.9077(51)(157)$ \\
Lattice error reduced to 1.8\%

\vspace{0.5in}
}
 Fig. 8 : Chiral/continuum extrapolation of 
the $B \rightarrow D^*$ form factor ${\cal F}(1)$.

\vspace{.1in}
\noindent
The fact that the central value of ${\cal F}(1)$ has come down between 2008 and 
2010 means that $|V_{cb}|_{excl.}$ has gone up.  At this meeting 
Mackenzie \cite{mackenzie} quotes $|V_{cb}|_{excl.} = 39.7(7)(7) \times 10^{-3}$ 
using Fermilab/MILC's new ${\cal F}(1)$ and an HFAG 09 experimental average 
for $|V_{cb}| * {\cal F}(1)$.  This updated $|V_{cb}|_{excl.}$ 
is to be compared with the PDG 2010 
$|V_{cb}|_{incl} = 41.5(7) \times 10^{-3}$.  The exclusive 
and inclusive determinations differ now by just $\sim1.5 \sigma$.
As has been pointed out by many authors, reducing uncertainty in 
$|V_{cb}|$ is of crucial importance in testing the SM via consistency 
checks on the Unitarity Triangle (UT).  Tensions have developed,
for instance, between
$\sin(2 \beta)$, $\epsilon_K$ and $\Delta M_s / \Delta M_d$.  
Analysis of the UT is very sensitive to $|V_{cb}|$. One has 
$\epsilon_K \propto |V_{cb}|^4$ and hence the uncertainty in $\epsilon_K$ is 
dominated by that of $|V_{cb}|$ (and not by errors in $B_K$).  
In order for $|V_{cb}|^4$ to have errors comparable to or better than 
that of $B_K$ (currently $\sim 4$\%) one would need to know 
$|V_{cb}|$ with an accuracy of $\leq 1$\%. The lattice community still 
has more work to do.

\vspace{.1in}
\noindent
Another important area in $B$ Physics where lattice input is crucial 
 is $B -\overline{B}$ mixing. This is a vast subject and Lattice QCD has 
so far addressed only parts of it.  In order to cover both SM and 
BSM physics even at lowest order in $1/M$ 
one needs hadronic matrix elements of a basis of 
five $\Delta B = 2$ four-fermion operators 
usually denoted $Q_i$, $i = 1,2 ...,5$.
Most of the published unquenched work has focused on $Q_1$, $Q_2$ and 
$Q_3$, operators relevant for the SM.  From these one can calculate 
bag parameters such as $\hat{B}_{B_q}$ ($q = s$ or $d$) and 
the important ratio $\xi = \frac{f_{B_s}\sqrt{B_{B_s}}}{f_{B_d} 
\sqrt{B_{B_d}}}$. $\xi$ can be combined with the experimentally 
measured mass differences $\Delta M_s$ and $\Delta M_d$ in 
$B_s - \overline{B}_s$ and $B_d - \overline{B}_d$ 
mixing to determine the ratio of CKM matrix elements 
$\frac{|V_{td}|}{|V_{ts}|}$.  
Fig. 9 shows some recent results for $\xi$ \cite{fermixi,hpqcdxi,rbcxi}. The HPQCD result 
is used to get $\frac{|V_{td}|}{|V_{ts}|}$.  
Their value for the bag parameter $\hat{B}_{B_s}$ also 
leads to the most precise SM prediction to date for the branching 
fraction $ Br(B_s \rightarrow \mu^+ \mu^-)$.

\parbox{0.5\hsize}{
\includegraphics*[height=12cm,angle=270]{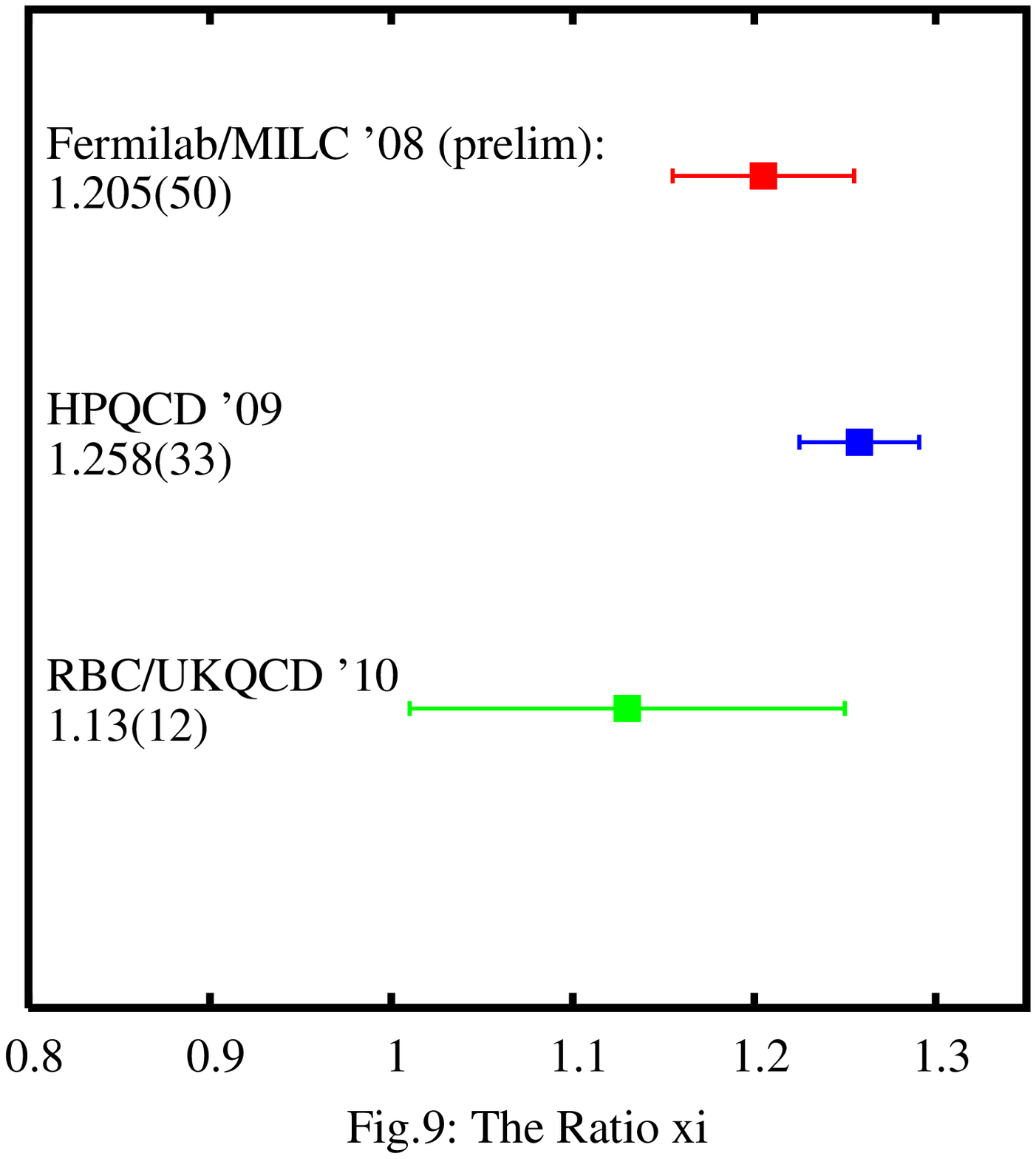}
}
\parbox{0.5\hsize}{

$\qquad$ HPQCD $\xi $ \cite{hpqcdxi} $\;\; \Longrightarrow$

\vspace{.1in}
$ \qquad \frac{|V_{td}|}{|V_{ts}|} = 0.214(1)(5)$

\vspace{.5in}
$ \qquad \hat{B}_{B_s} = 1.33(6) \; \; \Longrightarrow$

\vspace{.1in}
$\qquad  Br(B_s \rightarrow \mu^+ \mu^-) = 3.2(2) \times 10^{-9}$

\vspace{0.5in}
}

\vspace{.1in}
\noindent
Several lattice groups are currently working on improving 
calculations of four-fermion operator matrix elements.  There should be 
many results coming out soon including matrix elements of operators 
relevant for BSM physics. 

\noindent
Reference \cite{hpqcdxi} also has updates of $B$ meson decay constants 
$f_B$ and $f_{B_s}$.
  These are shown in Figs.10 and 11 together with 
preliminary results from the Fermilab/MILC \cite{fermifd10} and the ETM 
\cite{etmcfb} collaborations. 
All groups are working hard on reducing errors.  
  Accurate calculations of $f_B$ are important because of 
the current tension between lattice and experimental determinations 
(the latter depends critically on the value used for $|V_{ub}|$).

\noindent
During the past couple of years work has also been initiated on Rare B Becays. 
Some preliminary results from the Cambridge group 
for $B \rightarrow K$ form factors $f_+$, $f_0$ and $f_T$ and 
for the $B \rightarrow K^*$ form factors $T_1$ and $T_2$ were presented 
at this meeting \cite{liu}. 

\parbox{0.5\hsize}{
\includegraphics*[height=11.cm,angle=270]{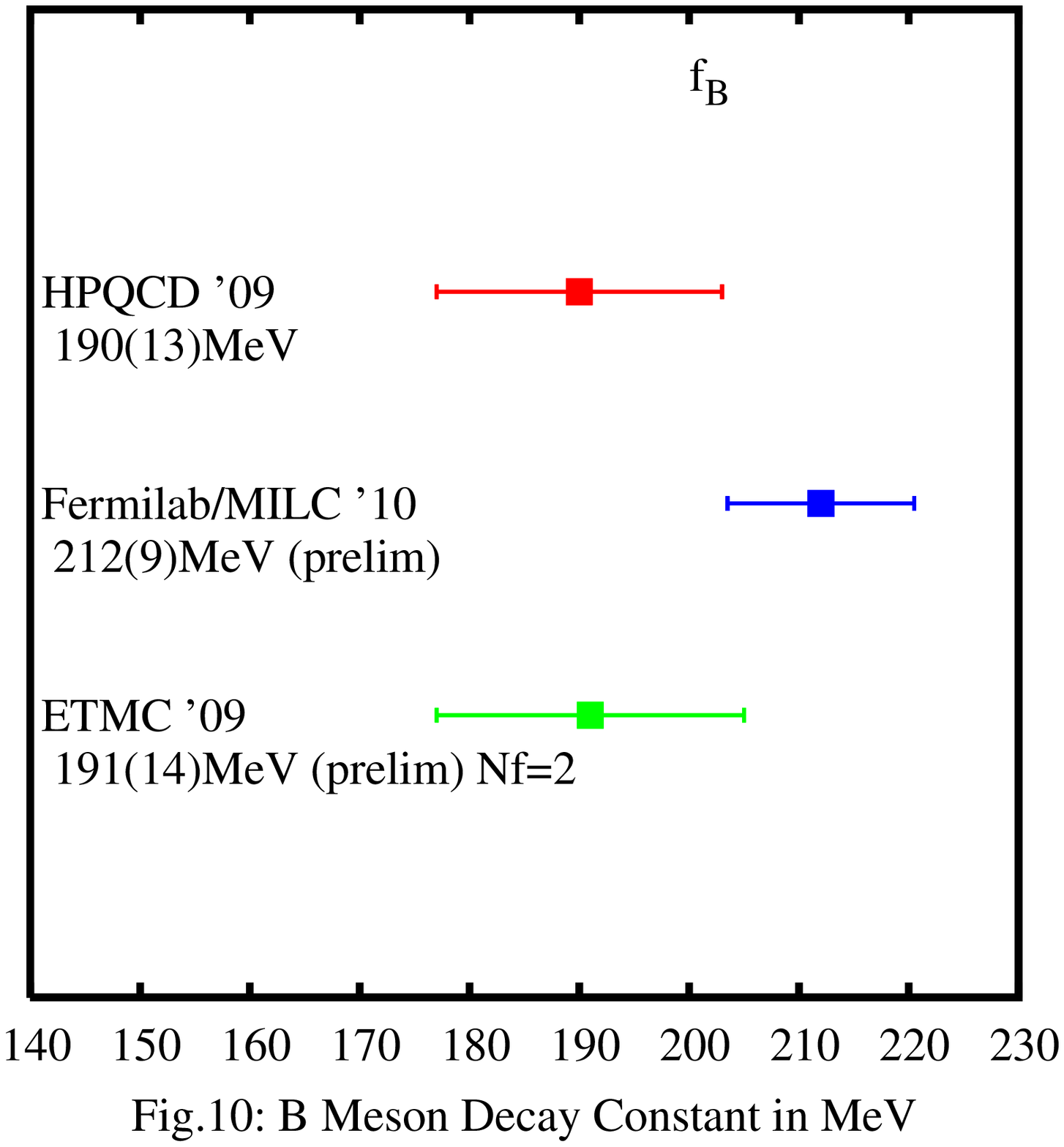}
}
\parbox{0.5\hsize}{
\includegraphics*[height=11.cm,angle=270]{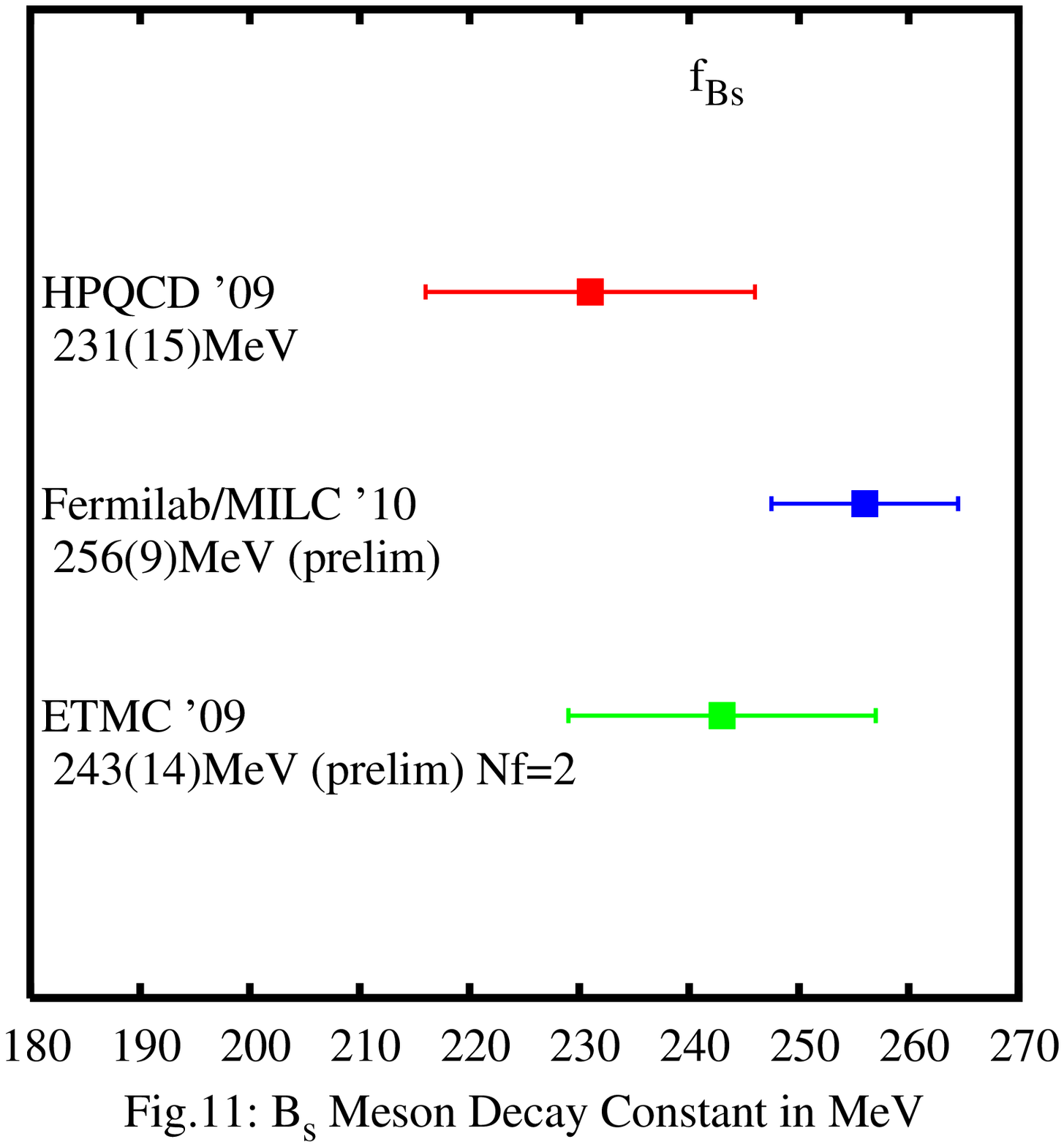}
}

%\vspace{.2in}

%\parbox{0.5\hsize}{
%\includegraphics*[height=7.cm]{B2K_fplus_f0_fT_q2.ps}
%}
%\parbox{0.5\hsize}{
%\includegraphics*[height=7.cm]{B2Kstar_T2_T1_q2.ps}
%}
%\hspace*{1.8in} Fig. 12 \hspace{2.in} Fig. 13 

\section{Summary}
It is hopefully clear by now that Lattice QCD is working hard to do its share 
in Precision Flavor Physics. More and more results with a few \% or better 
errors are becoming available making accurate tests of the Standard Model 
possible. We look forward to exciting times ahead as we continue to work 
together with our experimental and theory colleagues.

\Acknowledgements
I thank the organizers of CKM2010 for a very enjoyable and superbly run 
meeting.

\end{document}